# Continuum model for hydrogen pickup in Zirconium alloys of LWR fuel cladding

Xing Wang [a, #], Ming-Jie Zheng [b, c, §, #], Izabela Szlufarska [a, b], Dane Morgan [a, b, *]

[a] Department of Engineering Physics and Nuclear Engineering, University of Wisconsin, Madison, WI, 53706, USA

[b] Department of Materials Science and Engineering, University of Wisconsin, Madison, WI, 53706, USA

[c] Key Laboratory of Neutronics and Radiation Safety, Institute of Nuclear Energy Safety Technology, Chinese Academy of Sciences, Hefei, Anhui, 230031, China

*Corresponding author Email: ddmorgan@wisc.edu

[§] The work was done at the University of Wisconsin Madison.
[#] Equal contribution

**Abstract**

A continuum model for calculating the time-dependent hydrogen pickup fractions in various Zirconium alloys under steam and pressured water oxidation has been developed in this study. Using only one fitting parameter, the effective hydrogen gas partial pressure at the oxide surface, a qualitative agreement is obtained between the predicted and previously measured hydrogen pickup fractions. The calculation results therefore demonstrate that H diffusion through the dense oxide layer plays an important role in the hydrogen pickup process. The limitations and possible improvement of the model are also discussed.

## 1. Introduction

Zirconium (Zr) alloys have been widely used as cladding materials for nuclear fuels in light-water reactors (LWR). During the operation of the Zr cladded fuels, hydrogen generated by the water and Zr corrosion reactions permeates through the protective Zr oxide layers, diffuses and accumulates in Zr metal, potentially reaching or surpassing the hydrogen solubility limit in the Zr alloy [1]. This process is called hydrogen (H) pickup and it can lead to the formation of brittle hydrides that significantly reduce the ductility of Zr alloys [2]. Therefore H pickup is one of the major issues potentially limiting the reliability and durability of cladding materials, especially under high burnups and accident conditions such as loss-of-cooling accidents and reactivity-initiated accidents [3].

It is usually proposed that H pickup proceeds in three steps [4,5]. First, $H_2O$ molecules adsorbed at the oxide/water interface react with anion oxygen vacancies to leave protons ($H^+$) on the oxide surface. Some of the protons are discharged by electrons migrating from the oxide/metal interface and become H adsorbates ($H_{ad}$). Subsequently the $H_{ad}$ atom either reacts to form $H_2$ to be released as hydrogen gas or $H_{ad}$ is absorbed into the oxide. Second, the absorbed hydrogen atoms, possibly along with protons, migrate through the barrier oxide layer and reach the Zr metal surface. Third, due to the H concentration gradient, hydrogen diffuses into Zr metal and hydride precipitates out when the hydrogen concentration is high enough.

Many studies have focused on understanding the transport of hydrogen through the barrier oxide layer, as this transport is often regarded as the rate-limiting step for H pickup [5–7]. A number of factors, including oxide morphology, alloy additive elements and local stress, play important roles in the hydrogen transport process. It is widely accepted that the Zr oxide scale develops a double-layered structure during the oxidation [8]. The outer layer is formed by porous oxide with cracks and pores that provide fast ingress routes for hydrogen, while the inner layer consists of dense oxide and it is usually regarded as a diffusion barrier [8]. Recent transmission electron microscope (TEM) analysis found a suboxide ($Zr_3O$) region existing at the metal/oxide interface in some Zr alloys [9]. According to density functional theory (DFT) calculations, the hydrogen migration energy in the suboxide is higher than that in pure Zr, so the suboxide layer may also slow down the H diffusion and contribute to the diffusion barrier [10]. Using *in-situ* nuclear reaction analysis, Une *et al.* measured the deuterium concentration depth profile in oxide layers of Zr alloys corroded in $D_2O$ steam [7]. The result shows a nearly flat concentration profile in the outside layer followed by a steeply decreasing concentration in the inner layer, which agrees well with the anticipated higher diffusivity in the porous oxide and lower diffusivity in the dense oxide. After growing to certain thickness, the dense oxide layer typically become porous (a change referred to as the "transition") and the corrosion rate is suddenly increased [8,11]. Usually a new dense oxide layer will start growing after the transition so the oxidation of most Zr alloys show a periodic feature. Recent studies have discovered that H pickup process also follows the oxidation periodicity [12,13]. Besides the oxide morphology, additive elements (e.g., Fe, Cr, Nb, Sn and Ni) in different Zr alloys have substantial effects on the H pickup fraction. These elements may either behave like trapping sites and directly decrease the hydrogen diffusivity in the oxide [14,15], or they may form second phase precipitates that have been hypothesized to be a preferred path for hydrogen migration or a source for pores or cracks formation [16–18]. Due to the lattice mismatch between Zr oxide and metal, high compressive stress is generated in the oxide near the interface [19,20]. Raman spectroscopy measurement revealed that the stress varies cyclically and can be as large as several GPa [7]. DFT calculations found that under 1GPa compressive stress hydrogen diffusion coefficient in tetragonal $ZrO_2$ is only about 60% of the coefficient without stress at 600 K [15]. It is worth noting that there is also literature arguing that the diffusion of hydrogen through the barrier oxide layer is not the rate-limiting step [21,22]. Evidence from the chemical exchange experiments [17] and transmission electron microscope (TEM) analysis [23] suggests that micro pores/cracks exist even in the dense oxide layer. It has been proposed that hydrogen can penetrate to the oxide/metal interface via these flaws and the hydrogen cathodic reaction at the Zr metal surface is the rate-controlling process for H pickup. One technical difficulty in evaluating this hypothesis is that the observed crystallite boundary cracks or pores could also be formed by the TEM sample preparation process [8]. Therefore it is still an open question which process (or processes) is (are) rate-limiting for H pickup.

Previous studies have provided a number of insights into the detailed mechanism of H pickup and acquired a large body of data under various corrosion conditions. However, few theoretical models have been developed that take advantage of the accumulated knowledge to describe the overall H pickup process quantitatively. In particular, the accuracy for predicting H pickup has not been assessed for even what might be considered as the simplest model, which assumes only a rate limiting process of diffusion in the oxide barrier layer. In this paper we have therefore developed and assessed such a model. This type of a model is important as a baseline for more complex models that invoke additional phenomena, e.g., rate limiting surface reaction processes, second phase precipitation, crack and pore formation, strain effects, etc. Using up-to-date diffusivities and corrosion measurements, H pickup fractions ($f_H$) in six different alloys (Zry-2, GNF-Ziron, VB, Zry-4, ZIRLO and Zr-2.5Nb) were calculated and compared with the experimentally measured $f_H$ [7,12]. Here $f_H$ is defined as the fraction of hydrogen absorbed by the Zr metal to the total hydrogen generated during corrosion. $f_H$ has been widely used for comparing resistances of various Zr alloys to the H uptake under different corrosion conditions (e.g., temperature, corrosion solution) [7,12]. The alloys are chosen to include all alloys for which the necessary data on H diffusivity, time-depended weight gain, and experimentally-measured $f_H$ needed by the model for comparing reason are simultaneously available. In this study, corrosions in 360 °C pure water (for Zry-4, ZIRLO and Zr-2.5Nb), 400 °C steam and 290 °C LiOH-containing water (for Zry-2, GNF-Ziron and VB) were analyzed. In Ref. [12], the samples of Zry-4 and ZIRLO were processed in both sheet and tube form in order to test whether the sample geometry could affect the H pickup process. These data are also included and compared in this work. Previous investigations have confirmed that the H pickup rate is significantly accelerated by the LiOH addition to water [24]. Further TEM analysis discovered that extended networks of degraded grain boundaries were formed from the oxide surface to near the metal/oxide interface, probably due to the preferential dissolution of zirconia in LiOH solution [7]. In this case the H pickup process is controlled by the dissociation reaction of $H_2O$ at the front of the degraded grain boundaries, rather than the hydrogen diffusion process. Including the LiOH case aims at showing the limitation of current model and indicating possible improvement for future work.

## 2. Methods

In our model, hydrogen diffusion through the dense barrier oxide layer is taken as the rate-limiting step for H pickup [5–7]. The hydrogen diffusion follows the equation

$$\frac{\partial C_H^{ZrO_2}(x,t)}{\partial t} = \nabla \left( D_H \frac{\partial C_H^{ZrO_2}(x,t)}{\partial x} + D_H \frac{C_H^{ZrO_2}(x,t) q_H E}{k_B T} \right), \left( 0 \leq x \leq L_b(t) \right) \quad (1)$$

Here $C_H^{ZrO_2}(x,t)$ is the hydrogen concentration in the barrier oxide layer at distance $x$ to the oxide surface and time $t$. As shown on the right side of the equation, the H flux contains two terms. The first term is the flux from concentration gradient across the oxide film and the second term is due to the electric field generated by other migrating charge particles (e.g. oxygen ions, electrons) during the oxidation process. Previous studies suggest that at least part of hydrogen migrating through the oxide layer is charged [12,25], so it is necessary to include the effect of electric filed from other charged particles on the H pick up process. Here $D_H$ is the hydrogen diffusion coefficient. As mentioned earlier, alloy additives have substantial effects on the H diffusivity in Zr oxides and the measured $D_H$ for different Zr alloys can vary significantly. When multiple $D_H$ values are available for a single Zr alloy, we chose the $D_H$ value that is closest to the averaged $D_H$ values among all the Zr alloys. The chosen $D_H$ for the six Zr alloys being studied here are summarized in Table 1 [14,26–29]. The effect of different $D_H$ values on the final calculated H pickup fraction is analyzed in the discussion part. The electric field-induced H flux is calculated based on the steady state Nernst-Planck equation. In equation (1) $q_H$ is the charge of H ion (+1 unit charge), $k_B$ is Boltzmann constant and $T$ is corrosion temperature. $E$ represents the electric filed across the Zr oxide film and is calculated by

$$E = \rho j = \rho \left[ \frac{q_O N_A}{M_O} \frac{dwg(t)}{dt} \right] \quad (2)$$

Here $\rho$ is the electrical resistivity of the oxide (in MΩ•cm) and $j$ is the oxidation current density (in A/cm$^2$). According to previous studies, the values of $\rho$ vary as oxidation proceeds. Different Zr oxides can also have quite different $\rho$ values with a range of about 2 – 138 MΩ•cm [30–32]. In order to estimate the maximum possible effects of the electric field on the H pickup process, the maximum $\rho$ (138 MΩ•cm) among all the reported values in literature is chosen in our calculation. We assume the oxidation current density $j$ is proportional to the oxidation rate. In equation (2), $wg(t)$ is the time-depended weight gain of the Zr specimen. If the contribution of absorbed H to the weight gain is neglected, then $dwg(t)/dt$ is the oxidation rate (in mg/dm$^2 \cdot$s). $q_o$ is the charge of oxygen ion (+2 unit charge), $N_A$ is Avogadro's constant and $Mo$ is molar mass of oxygen ion (15.9994 g/mol). Our calculations show that the contribution of the electric filed to the H pickup is negligible when compared to the contribution of concentration gradient, since the oxidation current density $j$ decreases rapidly as the oxidation proceeds. The detailed analysis will be presented in the Discussion section.

In equation (1), $L_b(t)$ is the time-dependent thickness of the barrier oxide layer. In order to solve the equation, the value of $L_b(t)$ must be accessible at all times as an analytical function since Eq. (1) must be evaluated at different time steps for the numerical solution. Therefore, we need fit an analytical form for $L_b(t)$. It is difficult to directly measure the oxide thickness during the corrosion as the alloy specimen must be periodically taken out of the autoclave and analyzed, typically by electron microscopy. Instead, weight gains of the specimen as a function of corrosion time are usually reported in literature [12,21,22]. The weight gain can be related to the oxide thickness based on the overall corrosion reaction $Zr + 2H_2O \rightarrow ZrO_2 + 2H_2$. If assuming the weight gain arises only from the added oxygen and the oxide has a uniform thickness, the time-dependent oxide thickness $L(t)$ (including the protective barrier oxide layer $L_b(t)$ and the non-protective porous layer) can be calculated by

$$L(t) = \frac{M_{ZrO_2}}{M_{O_2}} \frac{wg(t)}{\rho_{ZrO_2}} \quad (3)$$

where $M_X$ is the molar mass of chemical X and $\rho_{ZrO2}$ is the zirconia density (5.68 g/cm$^3$). As the same in equation (2), $wg(t)$ is the time-depended weight gain of the Zr specimen. For diffusion controlled growth, the oxidation kinetics follows a power law yielding [33,34]

$$wg(t) = Kt^q \quad (4)$$

By fitting a series of weight gains measured at different time during the corrosion, the $K$ and $q$ can be obtained for all six alloys being studied here under steam or water corrosion. More specifically, the weight gain data for fitting $K$ and $q$ values of Zry-2, GNF-Ziron and VB is from Ref. [7] and the data for fitting $K$ and $q$ values of Zry-4, ZIRLO and Zr-2.5Nb is from Ref. [12]. For the corrosion in LiOH-containing water, as the surface reaction of H$_2$O and Zr is the controlling process, the $wg$ vs. time follows a simple linear relationship so $q$ is set equal to one and only $K$ is fitted [7]. The fitted coefficients are summarized in Table 1. $K$ has the same unit as $wg$ (mg/dm$^2$) and $q$ is a numerical factor corresponding to corrosion time in days. As mentioned earlier, the entire oxide layer ($L(t)$) contains both dense protective layer ($L_b(t)$) and porous non-protective layer. To determine $L_b$, nuclear reaction analysis was used to measure the deuterium concentration profiles in the oxide layer of Zry-2, GNF-Ziron and VB corroded in 400 °C $D_2O$ steam before the transition [7,14]. In the concentration profile, the region of a flat deuterium concentration is regarded as corresponding to the porous layer and the region of a decreasing concentration is regarded as being due to the dense protective oxide layer [7,14]. The measurement has found the thickness of the dense oxide layer is about 0.53-0.60 of the entire oxide layer. A large number of theoretical and experimental analyses suggests that the dense oxide undergoes a transition when it reaches its maximum thickness $L_{bm}$ [7–9,35,36]. Therefore in our calculation, it is assumed that before the transition, the barrier oxide layer grows with a constant thickness fraction ($\eta$) of the entire oxide layer, but after the transition, the dense oxide layer becomes porous and is no longer a barrier to H diffusion. We also assume that after the transition a new dense oxide layer starts growing, following the same kinetics as before the transition. Our assumption is consistent with the periodic feature of Zr oxidation and the fact that a thicker oxide layer is usually associated with a superior resistance to oxidation and H pickup [37]. With all these assumptions, we get $L_b(t)=\eta L(t)$ for $t<t_{transition}$. Here $t_{transition}$ is the time when the oxidation transition happens, which is indicated by a sudden increase of the oxidation rate shown in the measured weight gain-time curve [7,12]. The values of $\eta$ and $L_{bm}$ for Zry-2, GNF-Ziron and VB from Ref. [7] are listed in Table 1. For Zry-4, ZIRLO and Zr-2.5Nb, the values of $L_{bm}$ and $\eta$ are not reported and only the entire oxide thickness right before the transition was calculated in Ref. [12]. For these three alloys we assume that the fractions $\eta$ is equal to the average value of $\eta$ (0.57) of the other three Zr alloys in Ref. [7]. For corrosion in LiOH, the dense oxide layer is very thin or possibly does not exist at all [7,14]. However, in order to compare with the water corrosion case, we treat the entire oxide layer thickness $L(t)$ as a barrier layer for the H diffusion calculation. The failure of our model to explain the LiOH data both supports the model by showing it fails for systems where it does not include the correct physics and supports the hypothesis that little or no dense oxide exists in the LiOH system.

In order to solve equation (1), boundary conditions at the H₂O/oxide interface ($x=0$) and oxide/metal interface ($x=L_b(t)$) must be set. As the hydrogen diffusion is assumed to be the rate-limiting step, the hydrogen chemical potentials ($\mu_H$) on both sides of each interface are treated as equal. Under the equilibrium condition, the boundary conditions are

$$C_H^{ZrO_2}(x=0,t) = C_{H,1atm}^{ZrO_2}\sqrt{p_{H_2}} \tag{5}$$

$$C_H^{ZrO_2}(x=L_d,t) = \frac{\rho_{ZrO_2}}{\rho_{Zr}} C_H^{Zr}(t) \tag{6}$$

where $C_{H,1atm}^{ZrO_2}$ is the hydrogen solubility in zirconia at standard atmospheric pressure and $p_{H_2}$ is the effective H₂ partial pressure just outside the oxide. All concentrations in this work are given as mole fractions of the host unless stated otherwise. The equation (5) simply follows the classic Sievert's law and both equations (5) and (6) assume ideal-mixing behavior of the dissolved hydrogen in ZrO₂ and Zr, respectively. According to Ref. [38], $C_{H,1atm}^{ZrO_2}$ (unit: mol H/mol ZrO₂) equals to 2.78×10⁻⁴ at 400 °C, 3.80×10⁻⁴ at 360 °C and 7.29×10⁻⁴ at 290 °C. It is worth mentioning that here $p_{H_2}$ is only an effective pressure that represents the activity of H for entering the oxide. It is not meant to represent a real gas pressure at the water/oxide interface or the overall activity of H in the surrounding water. The $p_{H_2}$ represents the H activity that comes from the detailed reactions and H generation occurring right at the surface of the oxide. Currently no experimental data about the time-dependent $p_{H_2}$ or H activity at the oxide surface is available, so we hypothesize that $p_{H_2}$ is a single constant value during the entire corrosion process for all Zr alloys corroded by steam or pure water. The value of $p_{H_2}$ is fitted by minimizing the sum of the squares of the calculation error (defined as difference between the calculated $f_H$ and the measured $f_H$). The fitted $p_{H_2}$ is 3.35×10⁶ atm. Again, we note that this is an effective value representing local H activity at the interface and potentially weakly related or unrelated to H activity of the surrounding water. Furthermore, our calculations show that increasing or decreasing the $p_{H_2}$ value only makes the total calculation errors larger, but does not change the trend of the calculated $f_H$ on which our conclusions are based. Similar fitting has been performed for corrosion in LiOH and the fitted $p_{H_2}$ equals to $6.10 \times 10^7$ atm. In equation (6), $C_H^{Zr}$ is the time-dependent hydrogen concentration in Zr alloy and can be calculated by

$$C_H^{Zr}(t) = C_{H,0}^{Zr} + \frac{\rho_{Zr}}{d \times \rho_{ZrO_2}} \int_0^t D \frac{\partial C_H^{ZrO_2}}{\partial x}\bigg|_{x=L_b} dt \quad (7)$$

Here $d$ is the thickness of the Zr cladding, which is 600 μm for Zry-2, GNF-Ziron and VB [7], and 800 μm for the remaining three alloys according to the real sample size [12]. The first term in equation (7) is the intrinsic initial hydrogen concentration in the alloys and the second term represents the accumulated hydrogen due to the H flux from oxide into metal. Equation (7) assumes that no hydride precipitation occurs, which is consistent with the concentrations according to our model calculations. Based on Ref. [12,14], $C_{H,0}^{Zr}$ is 9 weight ppm for both Zry-2 and VB, 6 weight ppm for GNF-Ziron and about 12.5 ppm for the remaining three alloys. For initial conditions, we take $C_H^{ZrO_2} = C_{H,1atm}^{ZrO_2}\sqrt{p_{H_2}}$ for $x=0$ and $C_H^{ZrO_2} = \frac{\rho_{ZrO_2}}{\rho_{Zr}} C_{H,0}^{Zr}$ for $0 < x \leq L_b$, which equations fulfill the equilibrium condition for hydrogen chemical potentials at the water/oxide and oxide/alloy interface, respectively. With the boundary and initial conditions, the time evolution of the hydrogen concentration profile in zirconia is solved using the standard finite difference method implemented by us in Matlab. Based on the obtained hydrogen concentration in Zr, the H pickup fraction is calculated by

$$f_H^{cal}(t) = \frac{[C_H^{Zr}(t) - C_H^{Zr}(t=0)]}{wg_{exp}(t)} \times \frac{d\rho_{ZrO_2} m_{O_2}}{m_{ZrO_2}} \quad (8)$$

Here $wg_{exp}(t)$ is the experimentally measured weight gain at time $t$. Note that H pickup fraction in equation (8) is always calculated with respect to an experimentally measured weight gain. The power law expression in equation (4) for the weight gain is not used in equation. (8) and is only used to estimate the oxidation current density in equation (2) and the barrier layer thickness in equation (3). The time-depended $f_H$ calculated by our model is then compared with the experimentally measured $f_H$ values at the same time. In the experiments, the H concentrations in the Zr alloys is measured at a certain corrosion time by either the vacuum hot extraction method or cold neutron prompt gamma activation analysis. Subsequently the H concentrations are converted into the H pickup fractions with the weight gain measured at the same time [7,12].

Table 1. Input parameters for solving hydrogen diffusion in Zr oxide layer

| | alloy | $D$ (m$^2$/s) | $K$ (mg/dm$^2$) | $q$ | $L_{bm}$ (μm) | $\eta$ |
|---|---|---|---|---|---|---|
| **360°C water** | Zry-4 | $4.49\times10^{-19}$ [27] | 8.61 (s) | 0.29 (s) | 1.2 (s) | 0.57 |
| | | | 7.47 (t) | 0.33 (t) | 1.5 (t) | |
| | ZIRLO | $4.49\times10^{-19}$ * | 6.02 (s) | 0.41 (s) | 1.7 (s) | 057 |
| | | | 7.28 (t) | 0.37 (t) | 1.8 (t) | |
| | Zr-2.5 Nb | $1.81\times10^{-19}$ [29] | 7.15 | 0.38 | 2.0 | 0.57 |
| **400 °C steam** | Zry-2 | $3.37\times10^{-18}$ [26] | 9.81 | 0.38 | 1.6 | 0.60 |
| | GNF-Ziron | $2.16\times10^{-18}$ [14] | 11.85 | 0.32 | 1.5 | 0.53 |
| | VB | $8.90\times10^{-19}$ [14] | 9.28 | 0.34 | 1.8 | 0.57 |
| **290 °C LiOH** | Zry-2 | $1.50\times10^{-17}$ [28] | 16.26 | 1.00 | 1.4 | 1.00 |
| | GNF-Ziron | $1.08\times10^{-17}$ [28] | 17.81 | 1.00 | 1.4 | 1.00 |
| | VB | $1.19\times10^{-17}$ [28] | 18.71 | 1.00 | 1.4 | 1.00 |

*Note: no reliable data for H diffusivity in ZIRLO oxide is available that we could find, so the H diffusivity in Zry-4 oxide is used for ZIRLO because of similarities in composition (Zry-4: 1.45 Sn-0.2 Fe-0.1 Cr, ZIRLO: 1.0 Nb-1.0 Sn-0.1 Fe.). The experimental weight gain data for fitting K and q of Zry-2, GNF-Ziron and VB is from Ref. [7] and the data for fitting K and q values of Zry-4, ZIRLO and Zr-2.5Nb is from Ref. [12]. Since the Zry-4 and ZIRLO sample have both tube form and sheet form, and the oxidation kinetics are different between samples with different shapes, the values of K, q and $L_{bm}$ for both the sheet and tube samples of Zry-4 and ZIRLO are listed separately in the table. The "s" means the sample is in sheet form and "t" means the sample is in tube form.*

## 3. Results and Discussion

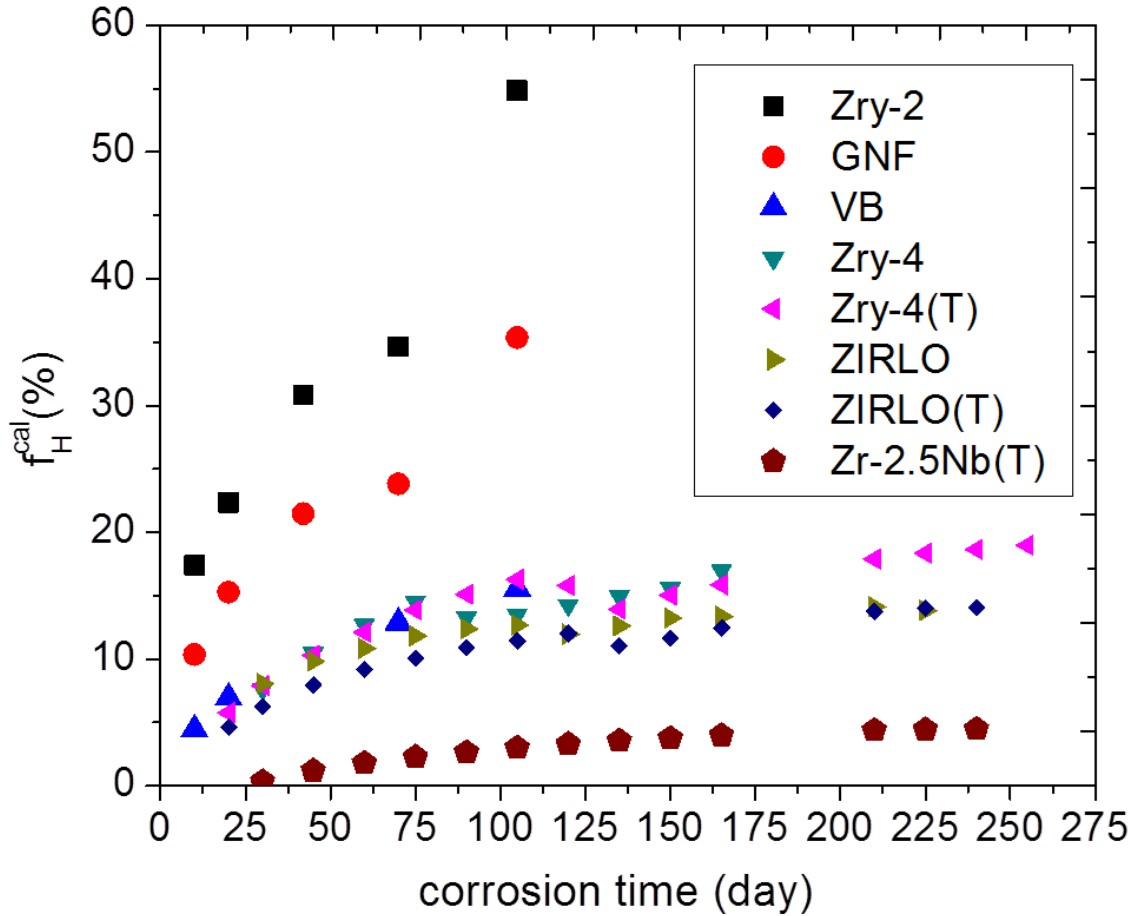

*Figure 1. Calculated $f_H$ vs. time under steam/water corrosion conditions. The letter (T) after the Zr alloy name means that the data was for the tube samples, otherwise the data was for the sheet samples.*

Figure 1 shows the calculated time-dependent H pickup fractions of all Zr alloys corroded in 400 °C steam or 360 °C water. In general, the pickup fractions increase with time, which trend agrees with the experiments. One exception is that for Zry-4 and ZIRLO, the $f_H$ has a slight but sharp decrease at the transition time (90 days for Zry-4 sheet, 120 days for ZIRLO sheet and 135 days for both Zry-4 tube and ZIRLO tube samples) and $f_H$ values for these alloys resume increasing afterwards. The sudden decrease is due to the fact that the oxidation rate, as well as the H generation rate, is suddenly accelerated at the oxidation transition. The same effect of the oxidation transition has also been observed in the experimentally measured H pickup fractions for these alloys, although the decrease is not as obvious as in the model and only a plateau of $f_H$ appears before the oxide transition time [12]. For Zry-2, GNF-Ziron, VB and Zr-2.5Nb, the increase of oxidation rate at the transition is not that pronounced so no similar $f_H$ decrease is shown in the calculations, and no decrease or plateau of $f_H$ was observed experimentally [7,12].

To directly compare the model calculations to the experimental measurements at the same corrosion time and conditions, the calculated fraction $f_H^{cal}$ vs. the experimentally measured fraction $f_H^{exp}$ are plotted in Figure 2a. The first letter in the symbol represents different alloys and the number is the corrosion time in days. The letter with prime symbol means the sample is in tube form, otherwise it is in sheet form. Since the data points of Zry-4, ZIRLO and Zr-2.5Nb are concentrated in the low $f_H$ corner, that region (marked by dotted line) is magnified in Figure 2b. According to Figure 2a and 2b, most of the data points are relatively close to the $f_H^{cal} = f_H^{exp}$ dashed line, which would correspond to a perfect agreement between the modeling predictions and experimental values. The values of the $f_H^{cal}$ and $f_H^{exp}$ shown in the Figure 2 are summarized in Appendix Table 1. According to the data, the average absolute error, which is defined as the average of absolute difference between $f_H^{cal}$ and $f_H^{exp}$ for all Zr alloys and at all different time, is 4.9%, and the average relative error, which is defined as the average of absolute difference between $f_H^{cal}$ and $f_H^{exp}$ divided by $f_H^{exp}$ for all Zr alloys and at all different time, is 61.0%. Given the uncertainties in the experimental measurements and the input parameters (e.g. H diffusivity and dense oxide layer thickness), as well as the various simplifying assumptions in the model, the agreement between $f_H^{cal}$ and $f_H^{exp}$ is relatively good. This agreement shows that the dense oxide layer plays a significant role in the H pickup process, and suggests that hydrogen diffusion through the dense oxide barrier layer may be the rate-limiting process for H pickup in many situations.

As an example of how the model can provide insights on what mechanisms are dominant for H pickup, our calculations show that the effect of electric filed on the H pickup is negligible when compared to the effect of the H concentration gradient across the oxide layer. To demonstrate this point, we calculated the H pickup fractions without considering the electric filed-induced H flux (i.e. the second term on the right side of equation (1)). The H pickup fractions with the effect of the electric filed (i.e. $f_{H,elec}^{cal}$) and without the effect ($f_{H,non-elec}^{cal}$) are also summarized in Appendix Table 1. The average relative difference between $f_{H,elec}^{cal}$ and $f_{H,non-elec}^{cal}$, which is defined as the average of the absolute differences between $f_{H,elec}^{cal}$ and $f_{H,non-elec}^{cal}$ divided by $f_{H,elec}^{cal}$ for all six types of Zr alloys and at all different corrosion time, is only 6.9%. It is worth noting that in order to estimate the maximum possible effect of the electric field, the maximum oxide resistivity we found measured in the literature (138 MΩ•cm) has been used in the calculation. If the average value of the oxide resistivity (70 MΩ•cm) is used, the average difference between $f_{H,elec}^{cal}$ and $f_{H,non-elec}^{cal}$ will be only 3.5%. The effect of the electric field on H pickup is significant at early times, e.g. within the first hour or so (in this paper all times are relative to the start of oxidation), but becomes very small when considered over the more relevant longer time scales of many days. The effect of electric filed on the H pickup process is negligible over long times because the oxidation current density $j$, as well as the associated electric filed $E$, decreases rapidly to low values as the oxidation proceeds, since $j$ is proportional to the time derivative of weight gain, which evolves with a negative exponent ($q-1$) of time. For example, $j$ in Zry-4 oxide is 5.44×10$^{-5}$ A/cm$^2$ at 30 mins, 3.32×10$^{-5}$ A/cm$^2$ at 1 hour and just 3.48×10$^{-6}$ A/cm$^2$ at 1 day. Correspondingly, the ratio of electric filed-induced H flux over the concentration gradient-induced H flux is 17.7 at 30 minutes, 0.16 at 1 hour and only about 0.02 at 1 day. This ratio remains lower than 2% in the rest of the corrosion process.

A careful examination of the distribution of the data points can reveal a few systematic discrepancies between the experimental and calculated $f_H$. First, as shown in Figure 2, the calculated H pickup fractions are higher than experimental values for most data points of Zry-4 and ZIRLO (both sheet and tube samples), whereas lower for Zr-2.5Nb, GNF-Ziron and VB. Second, as shown in Figure 2a, the $f_H^{cal}$ of Zry-2 is lower than $f_H^{exp}$ for shorter corrosion time but higher than $f_H^{exp}$ for longer corrosion time. This means the slopes of $f_H$ with time are steeper in the model calculations than in the experiments. In fact a similar error in the slope of $f_H$ with time also exists for GNF-Ziron and VB. Third, some experiments showed that the speed of H pickup suddenly increases just before the end of the oxidation transition [12,39]. However, such a phenomenon is not obviously observed in our calculated H pickup fractions.

Several possible reasons may contribute to the above discrepancies. First, the uncertainty in the measured H diffusion coefficients in various Zr alloy oxides is relatively large, even in the same Zr alloy. For example, according to Ref. [29], $D_H$ = $1.81 \times 10^{-19}$ m$^2$/s for Zr-2.5Nb at 360 °C, whereas in Ref. [40] $D_H$ is as large as $1.13 \times 10^{-17}$ m$^2$/s. This large difference may be caused by different techniques for measuring H concentration, details of sample preparations and methods for deriving the diffusion coefficient. The large uncertainties of $D_H$ can substantially influence the calculated H pickup fractions. A detailed estimation of the uncertainties of $D_H$ and other input parameters, as well as their effects on the calculated H pickup fractions is presented in the next paragraph. In addition, in our model a fixed $D_H$ is applied for calculating $f_H$ in the entire H pickup process. This fixed $D_H$ is usually derived by fitting the H concentration profiles measured at certain period of time after the corrosion starts [29,41,42]. Therefore the obtained $D_H$ is actually "averaged effective" diffusivity, while the real diffusivity in different regions of the oxide barrier layer or at different times may not be necessarily the same. For example, as mentioned in the introduction, the compressive stress in the oxide can decreases the H diffusivity. It is possible that during the initial oxidation, the stress in the thin oxide layer is still small so that the real H diffusivity is actually larger than the effective diffusivity, which would lead to a higher H pickup rate at short corrosion time. Second, an accurate model for the oxide growth is missing in our calculation. Here we assume the oxidation rate is proportional to the speed of weight gain and the weight gain follows the simple power law in equation (4). However, the growth of the oxide layer may not necessarily follow this simple kinetics, especially during the oxidation transition. Third, due to the difficulty in measuring H activity at the corroding surface during the corrosion test, a constant effective H$_2$ partial pressure $p_{H_2}$ is used for all the $f_H$ calculation. Here the partial pressure $p_{H_2}$ is simply representing the activity of H to enter the oxide. As the rate of corrosion reaction changes during the H pickup process, it is quite possible that $p_{H_2}$ also varies with time. Finally, as mentioned in the Introduction section, some other physical mechanisms that are not included in the current model may also substantially affect the H pickup process. For example, the formation of pores/cracks may provide fast ingress route for H, and the water splitting and H cathodic reactions on the Zr surface may also be a rate-limiting step for H pickup [17,23]. Furthermore, second phase precipitates can act as short-circuit paths for H through the dense oxide layer [11]. Given all these uncertainties in parameters and assumptions in oxidation kinetics, we do not claim that the current model can give quantitatively accurate predictions of H pickup fractions. However, this model provides useful qualitative guidance and acts as a baseline for more complex models with more accurate diffusion parameters, oxidation kinetics and physical mechanisms.

Here we analyzed the uncertainty of the model input parameters, including H diffusion coefficient, dense oxide layer thickness and $H_2$ partial pressure on the oxide surface, and their effects on the calculated H pickup fractions. For the H diffusion coefficient, as summarized in Table 1, the $D_H$ values chosen in our calculations are between $1.0 \times 10^{-18}$ $m^2/s$ and $1.0 \times 10^{-19}$ $m^2/s$ (at $T=360$ °C). However, the total range reported for $D_H$ is somewhat larger (a summary of all H diffusion coefficients reported by previous literature for various Zr alloy oxides can be found in Appendix Table 2). Based on the values in Appendix Table 2, we can calculate the standard deviation $\sigma$ of $\ln(D_H)$ and the mean $\lambda$ of $\ln(D_H)$. If we take $D_{H, Max}=\exp(\lambda+\sigma)$ and $D_{H, Min}=\exp(\lambda-\sigma)$, the range of $D_H$ is ($3.14 \times 10^{-20}$ $m^2/s$, $7.06 \times 10^{-18}$ $m^2/s$) at $T=360$ °C and ($6.27 \times 10^{-20}$ $m^2/s$, $1.13 \times 10^{-17}$ $m^2/s$) at $T=400$ °C. This large uncertainty of $D_H$ can substantially affect the calculation results. If $D_{H, Max}$ is used, the $f_H^{cal}$ for all Zr alloys will quickly increase to nearly 100% within about 40 days. If $D_{H, Min}$ is used, the $f_H^{cal}$ for all Zr alloys will remain nearly 0% during the entire corrosion process. Under these extreme cases, our model cannot give a reasonable prediction of the H pickup fractions. Although we chose the $D_H$ values in Table 1 following a logical approach (described in the Method section), the large uncertainty of $D_H$ and its substantial effects on $f_H^{cal}$ suggest that we should regard the $D_H$ as a partially fitted parameter. For the dense oxide layer thickness, previous studies show that the ratio of the dense oxide layer to the total oxide layer ranges from $\eta = 0.53$ to $0.60$ [7]. By using this range we can estimate that the potential uncertainty in the thickness of the dense oxide layer is within 0.2 μm for all kinds of Zr alloys studied here. This variance in the oxide layer thickness has relatively small effects on the $f_H^{cal}$. More specially, if the maximum oxide thickness is used (based on $\eta = 0.60$), the average absolute error between the $f_H^{cal}$ and $f_H^{exp}$ is 4.6% and the average relative error is 56.2%. If the minimum oxide thickness is used (based on $\eta = 0.53$), the average absolute error is 5.3% and the average relative error is 68.4%. Both errors in each case are very similar to the original calculation results, where the absolute error is 4.9% and the relative error is 61.0%. The calculated H pickup fractions using different oxide thickness are listed in Appendix Table 3. For the effective $H_2$ partial pressure $p_{H_2}$, a ten-fold cross validation was performed to evaluate its uncertainty range and the effects on the H pickup calculations. The original data set was partitioned into ten groups. Each time nine groups were selected as the training set to get the fitted $p_{H_2}$ using the same procedures as described in the Method section, then the fitted $p_{H_2}$ was applied to calculate the H pickup fractions in the final single group, which is called the validation set. This process was repeated ten times so each of the ten groups had been used as the validation set once. The detailed calculation results are summarized in the Appendix Table 4. The range of the fitted $p_{H_2}$ is between $2.6 \times 10^6$ atm to $3.7 \times 10^6$ atm, which are quite close to the original value of $p_{H_2} = 3.35 \times 10^6$ atm. The average absolute error of the validation set is 5.3% and the average relative error is 65.1%, which are also close to the corresponding values in the original calculations. We also plotted the $f_H^{cal}$ vs. $f_H^{exp}$ of the validation set in Appendix Figure 1. Most of the data points are relatively close to the $f_H^{cal} = f_H^{exp}$ dashed line, which is also similar to Figure 2a. Therefore the cross validation shows the $p_{H_2}$ is well constrained by the data and the likely errors from fitting have limited influence on the H pickup fractions. In summary, the H diffusion coefficient seems to affect the calculations most significantly. The large uncertainty rage of $D_H$ suggests that more high-quality diffusion data is necessary for fully assessing the accuracy of our model.

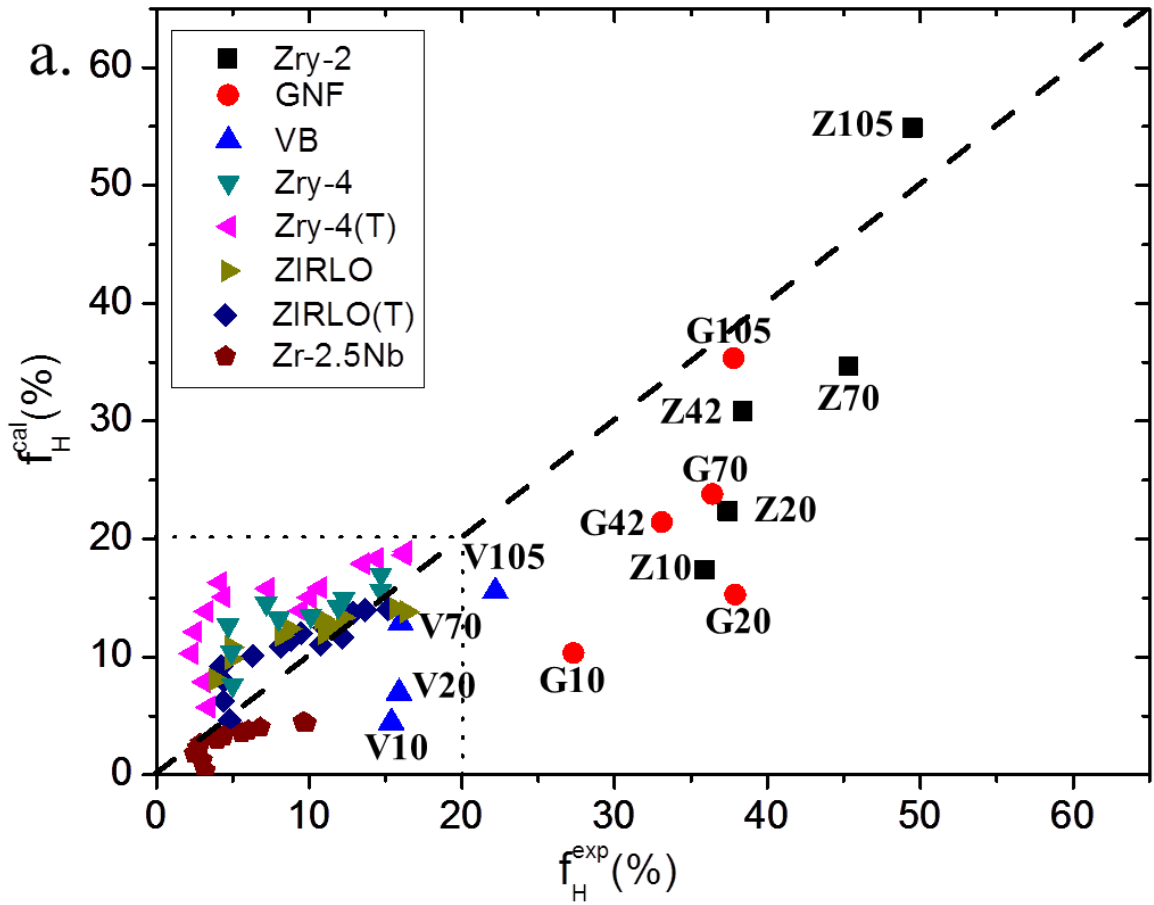

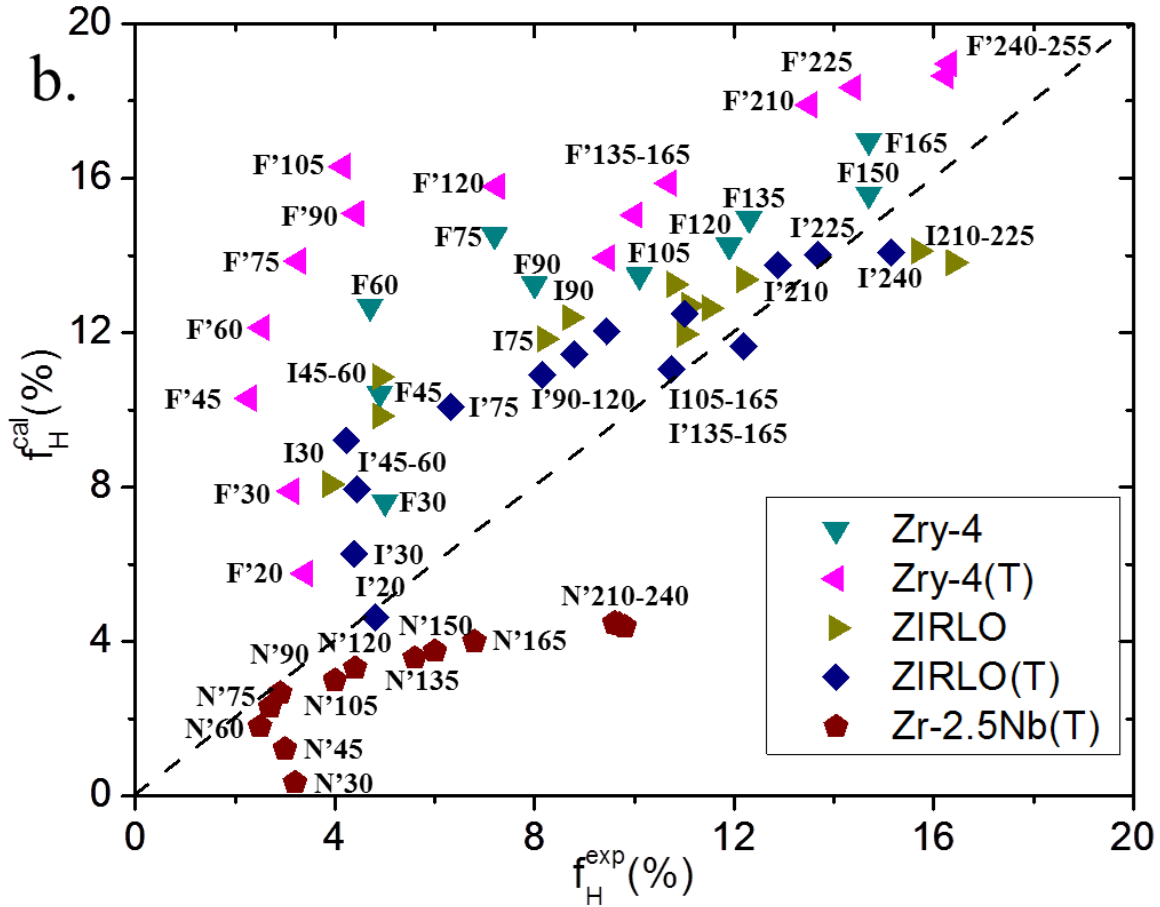

*Figure 2. (a) Comparison between the $f_H$ calculated by our model and measured experimentally under steam/water corrosion. For the name of each data point, the first letter represents the type of the Zr alloy and the number is the corrosion time in days. Z, G, and V stand for Zry-2, GNF-Ziron and VB, respectively. (b) Magnification of the left-bottom corner of Figure 2a. F, I, and N stand for Zry-4, ZIRLO and Zr-2.5Nb, respectively. The letter with prime symbol indicates that the sample is in tube form. The experimentally measured $f_H$ values of Zry-2, GNF-Ziron and VB are from Ref. [7] and the values of Zry-4, ZIRLO and Zr-2.5Nb are from Ref. [12] . The dashed line in both figures indicates the condition that calculated $f_H$ is equal to the experimentally measured $f_H$.*

To further validate the model we demonstrate that it fails where it is expected not to work. In Figure 3 we plot the calculated H pick up fraction $f_H^{cal}$ as a function of the experimentally measured fraction $f_H^{exp}$ for LiOH solution corrosion experiments. The detailed calculation results are listed in Appendix Table 5. Here the model is expected to show poor correlation to experiment, as no dense oxide layer is believed to form on Zr surface in the LiOH solution [7]. As expected, most calculated H pickup fractions are far from the experimental values, which demonstrates that when diffusion through the dense oxide is not a rate limiting process, our model does not predict reasonable $f_H$ values. More quantitatively, the average absolute error is 23.1% and the average relative error is 138.5% in the LiOH solution case. Both errors are much larger than the corresponding errors in the water/steam case. The failure of the model when applied to the LiOH solution case (where other physical phenomena are expected to affect H transport) supports the assertion that the model captures real physics when it successfully matches experimental data in the steam/water corroded materials. In addition, the success of the model for the steam/water corroded systems further supports the notion that the dense oxide plays a significant role in controlling $f_H$ in those measurements.

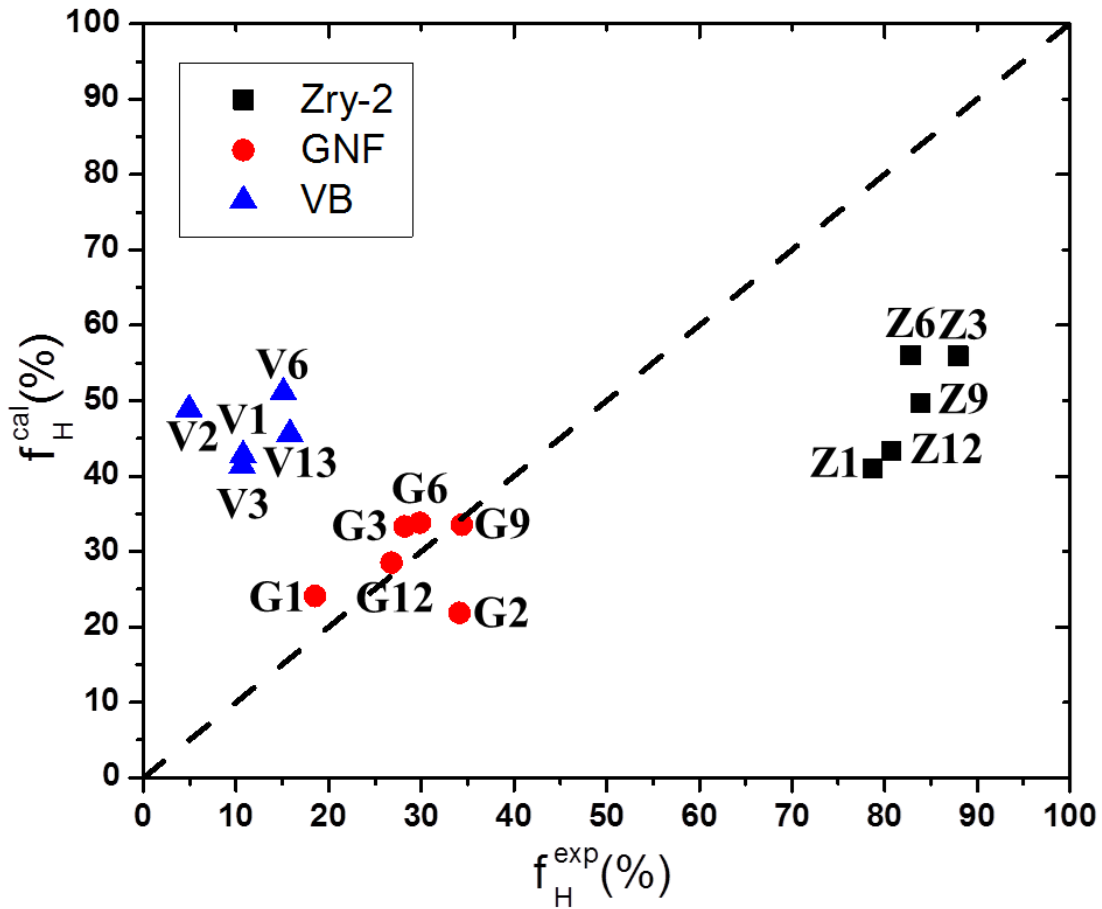

*Figure 3. Comparison between the $f_H$ calculated by our model and measured experimentally in 290 °C LiOH water. For the name of each data point, the first letter represents the type of the Zr alloy and the number is the corrosion time in days. The dashed line indicates the condition where the calculated $f_H$ is equal to the experimentally measured $f_H$. The experimentally measured $f_H$ values of Zry-2, GNF-Ziron and VB are from ref.[7].*

## 4. Conclusions

A continuum model for calculating the time-dependent H pickup fractions in different Zr alloys has been developed in this paper. To the best of our knowledge, all available experimental data that simultaneously measures time-depended H pickup fractions and weight gain for Zr alloys of available H diffusivities in the oxide layer have been collected and compared with the $f_H$ calculated by our model. The model's predictions match qualitatively with the experimental values of steam/water corrosion, which supports the hypothesis that hydrogen diffusion through the dense oxide barrier layer plays a significant role in affecting the H pickup process. The model breaks down when attempting to predict $f_H$ in LiOH containing water corrosion, providing further evidence for the absence of dense oxide in the LiOH environment and supporting the model through demonstrating that it fails when expected. This model offers a primary framework for developing more sophisticated models in the future when more accurate parameters (e.g., H diffusivity and solubility, oxide barrier layer thickness and effective $H_2$ partial pressure) are available, and for incorporating more physical mechanisms that may play an important role in H transport (e.g., pores/cracks formation, water dissociation on oxide surface, second phase precipitation and strain effects).

**Acknowledgements**

This research was primarily supported by the Consortium for Advanced Simulation of Light Water Reactors (http://www.casl.gov), an Energy Innovation Hub (http://www.energy.gov/hubs) for Modeling and Simulation of Nuclear Reactors under U.S. Department of Energy Contract No. DE-AC05-00OR22725. Support for Xing Wang to collect additional data, perform analysis, refine the codes, and write parts of the manuscript was provided by the US Department of Energy, Office of Basic Energy Sciences (grant DE-FG02-08ER46493). We are very grateful to Adrien Couet for helpful discussions.

# Appendix of Supplementary Data

*Appendix Table 1. Compare the H pickup fractions measured experimentally ($f_H^{exp}$), H pickup fractions calculated with effects of concentration gradient and electric filed ($f_{H,elec}^{cal}$), and H pickup fractions calculated with effects of concentration gradient only ($f_{H,non-elec}^{cal}$)*

| Alloy | Corrosion time (day) | $f_H^{exp}$(%) | $f_{H, elec}^{cal}$(%) | $f_{H, non-elec}^{cal}$(%) |
|---|---|---|---|---|
| Zry-2 | 10 | 35.9 | 17.39 | 21.32 |
| | 20 | 37.4 | 22.36 | 26.34 |
| | 42 | 38.4 | 30.85 | 35.74 |
| | 70 | 45.3 | 34.64 | 39.69 |
| | 105 | 49.5 | 54.86 | 62.19 |
| GNF-Ziron | 10 | 27.3 | 10.34 | 10.64 |
| | 20 | 37.9 | 15.28 | 15.61 |
| | 42 | 33.1 | 21.46 | 21.85 |
| | 70 | 36.4 | 23.83 | 24.23 |
| | 105 | 37.8 | 35.34 | 35.88 |
| VB | 10 | 15.4 | 4.47 | 5.17 |
| | 20 | 15.9 | 6.95 | 7.68 |
| | 70 | 16.0 | 12.87 | 13.67 |
| | 105 | 22.2 | 15.56 | 16.39 |
| Zry-4(sheet) | 30 | 5.0 | 7.62 | 8.43 |
| | 45 | 4.9 | 10.44 | 11.28 |
| | 60 | 4.7 | 12.68 | 13.54 |
| | 75 | 7.2 | 14.53 | 15.40 |
| | 90 | 8.0 | 13.26 | 14.17 |
| | 105 | 10.1 | 13.49 | 14.37 |
| | 120 | 11.9 | 14.26 | 15.14 |
| | 135 | 12.3 | 14.95 | 15.82 |
| | 150 | 14.7 | 15.58 | 16.45 |
| | 165 | 14.7 | 16.98 | 17.97 |
| Zry-4(tube) | 20 | 3.4 | 5.76 | 6.57 |
| | 30 | 3.1 | 7.89 | 8.73 |
| | 45 | 2.3 | 10.29 | 11.16 |
| | 60 | 2.5 | 12.12 | 13.00 |
| | 75 | 3.3 | 13.85 | 14.76 |
| | 90 | 4.4 | 15.09 | 16.00 |
| | 105 | 4.2 | 16.30 | 17.22 |
| | 120 | 7.2 | 15.79 | 16.63 |
| | 135 | 9.4 | 13.94 | 14.75 |
| | 150 | 10.0 | 15.05 | 15.94 |
| | 165 | 10.7 | 15.87 | 16.79 |
| | 210 | 13.5 | 17.90 | 18.85 |
| | 225 | 14.4 | 18.34 | 19.29 |
| | 240 | 16.2 | 18.65 | 19.59 |
| | 255 | 16.3 | 18.96 | 19.90 |
| ZIRLO(sheet) | 30 | 3.9 | 8.07 | 8.97 |
| | 45 | 4.9 | 9.84 | 10.77 |
| | 60 | 4.9 | 10.85 | 11.78 |
| | 75 | 8.2 | 11.83 | 12.78 |
| | 90 | 8.7 | 12.39 | 13.33 |
| | 105 | 11.1 | 12.69 | 13.62 |
| | 120 | 11.0 | 11.96 | 12.88 |
| | 135 | 11.5 | 12.63 | 13.58 |
| | 150 | 10.8 | 13.24 | 14.22 |
| | 165 | 12.2 | 13.37 | 14.33 |
| | 210 | 15.7 | 14.11 | 15.07 |
| | 225 | 16.4 | 13.81 | 14.74 |
| ZIRLO(tube) | 20 | 4.8 | 4.64 | 5.47 |
| | 30 | 4.4 | 6.27 | 7.12 |
| | 45 | 4.4 | 7.95 | 8.82 |
| | 60 | 4.2 | 9.21 | 10.09 |
| | 75 | 6.3 | 10.08 | 10.96 |
| | 90 | 8.2 | 10.92 | 11.81 |
| | 105 | 8.8 | 11.44 | 12.32 |
| | 120 | 9.4 | 12.04 | 12.92 |
| | 135 | 10.7 | 11.06 | 11.83 |
| | 150 | 12.2 | 11.65 | 12.56 |
| | 165 | 11.0 | 12.49 | 13.45 |
| | 210 | 12.9 | 13.76 | 14.74 |
| | 225 | 13.7 | 14.02 | 15.00 |
| | 240 | 15.2 | 14.07 | 15.04 |
| Zr-2.5Nb | 30 | 3.2 | 0.35 | 0.37 |
| | 45 | 3.0 | 1.22 | 1.24 |
| | 60 | 2.5 | 1.80 | 1.82 |
| | 75 | 2.7 | 2.30 | 2.32 |
| | 90 | 2.9 | 2.66 | 2.68 |
| | 105 | 4.0 | 3.00 | 3.02 |
| | 120 | 4.4 | 3.31 | 3.33 |

| | | | |
|---|---|---|---|
| 135 | 5.6 | 3.58 | 3.60 |
| 150 | 6.0 | 3.76 | 3.78 |
| 165 | 6.8 | 4.01 | 4.03 |
| 210 | 9.8 | 4.39 | 4.42 |
| 225 | 9.7 | 4.45 | 4.48 |
| 240 | 9.6 | 4.49 | 4.51 |

*Appendix Table 2. Summary of H diffusion coefficient in oxide of different Zr alloys (NRA is nuclear reaction analysis, GRA is gas release analysis and SIMS is second ion mass spectroscopy analysis)*

| Zr alloy | Diffusion prefactor (m$^2$/s) | Activation Energy (kJ/mol) | Investigator | Method |
|---|---|---|---|---|
| Zry-2 | $2.76 \times 10^{-9}$ | 114.84 | Khatamian [26] | NRA |
| Zry-2 | $1.30 \times 10^{-13}$ | 81.1 | Kunz [43] | GRA |
| Zry-2 | $4.00 \times 10^{-17}$ | 30.1 | Austin [44] | GRA |
| GNF-Ziron | $4.50 \times 10^{-17}$ | 17 | Takagi [14] | NRA |
| VB | $8.9 \times 10^{-19}$ at 673K | - | Une [7] | NRA |
| Zry-4 | $2 \times 10^{-21}$ (300K); $6 \times 10^{-19}$ (673K) | - | Hatano [27] | SIMS |
| Zr-2.5 Nb | $8.09 \times 10^{-18}$ | 20 | McIntyre [29] | SIMS |
| Zr-2.5 Nb | $3.05 \times 10^{-13}$ | 53.7 | Khatamian [42] | NRA |
| Zr-2.5 Nb | $1.15 \times 10^{-10}$ | 71.6 | Khatamian [40] | NRA |
| Zr-2.5Nb | $2.7 \times 10^{-19}$ (523K); $6.5 \times 10^{-19}$ (573K) | - | Une [45] | NRA |
| Zr-2.5Nb | $1 \times 10^{-18}$ (573 K) | - | Elmoselhi [29] | SIMS |
| Zr-20Nb | $2.60 \times 10^{-6}$ | 149.92 | Khatamian [26] | NRA |
| Zr-20Nb | $1.64 \times 10^{-8}$ | 118.7 | Urbanic [46] | NRA |
| Zr-15Nb | $1.99 \times 10^{-10}$ | 89.46 | Khatamian [26] | NRA |
| Zr | $1.13 \times 10^{-12}$ | 81.1 | Khatamian [42] | NRA |

*Appendix Table 3. Compare the calculated H pickup fractions using the maximum dense oxide layer thickness ($f_{H,max,oxide}^{cal}$), and the minmum dense oxide layer thickness ($f_{H,min,oxide}^{cal}$)*

| Alloy | Corrosion time (day) | $f_{H,\,max,oxide}^{cal}$ (%) | $f_{H,\,min,oxide}^{cal}$ (%) |
|---|---|---|---|
| Zry-2 | 10 | 17.4 | 18.8 |
| | 20 | 22.4 | 24.1 |
| | 42 | 30.8 | 33.1 |
| | 70 | 34.6 | 37.2 |
| | 105 | 54.9 | 58.7 |
| GNF-Ziron | 10 | 9.7 | 10.3 |
| | 20 | 14.5 | 15.3 |
| | 42 | 20.4 | 21.5 |
| | 70 | 22.7 | 23.8 |
| | 105 | 33.6 | 35.3 |
| VB | 10 | 4.1 | 5.0 |
| | 20 | 6.5 | 7.6 |
| | 70 | 12.1 | 13.9 |
| | 105 | 14.7 | 16.7 |
| Zry-4(sheet) | 30 | 7.0 | 8.5 |
| | 45 | 9.7 | 11.4 |
| | 60 | 11.9 | 13.8 |
| | 75 | 13.6 | 15.8 |
| | 90 | 12.4 | 14.4 |
| | 105 | 12.7 | 14.6 |
| | 120 | 13.4 | 15.4 |
| | 135 | 14.1 | 16.2 |
| | 150 | 14.7 | 16.8 |
| | 165 | 16.0 | 18.3 |
| Zry-4(tube) | 20 | 5.2 | 6.5 |
| | 30 | 7.3 | 8.8 |
| | 45 | 9.6 | 11.3 |
| | 60 | 11.3 | 13.2 |
| | 75 | 13.0 | 15.0 |
| | 90 | 14.2 | 16.3 |
| | 105 | 15.3 | 17.6 |
| | 120 | 14.9 | 17.0 |
| | 135 | 13.1 | 15.1 |
| | 150 | 14.2 | 16.2 |
| | 165 | 15.0 | 17.1 |
| | 210 | 16.9 | 19.3 |
| | 225 | 17.3 | 19.7 |
| | 240 | 17.6 | 20.1 |
| | 255 | 17.9 | 20.4 |
| ZIRLO(sheet) | 30 | 7.4 | 9.0 |
| | 45 | 9.1 | 10.8 |
| | 60 | 10.1 | 11.9 |
| | 75 | 11.1 | 12.9 |
| | 90 | 11.6 | 13.5 |
| | 105 | 11.9 | 13.8 |
| | 120 | 11.2 | 13.0 |
| | 135 | 11.9 | 13.7 |
| | 150 | 12.4 | 14.4 |
| | 165 | 12.6 | 14.5 |
| | 210 | 13.3 | 15.3 |
| | 225 | 13.0 | 14.9 |
| ZIRLO(tube) | 20 | 4.1 | 5.3 |
| | 30 | 5.7 | 7.0 |
| | 45 | 7.3 | 8.8 |
| | 60 | 8.5 | 10.1 |
| | 75 | 9.4 | 11.0 |
| | 90 | 10.2 | 11.9 |
| | 105 | 10.7 | 12.4 |
| | 120 | 11.3 | 13.1 |
| | 135 | 10.4 | 12.0 |
| | 150 | 10.9 | 12.7 |
| | 165 | 11.7 | 13.6 |
| | 210 | 12.9 | 14.9 |
| | 225 | 13.2 | 15.2 |
| | 240 | 13.2 | 15.2 |
| Zr-2.5Nb | 30 | 0.1 | 0.7 |
| | 45 | 0.9 | 1.6 |
| | 60 | 1.5 | 2.2 |
| | 75 | 2.0 | 2.7 |
| | 90 | 2.3 | 3.1 |
| | 105 | 2.7 | 3.5 |
| | 120 | 3.0 | 3.8 |
| | 135 | 3.2 | 4.1 |
| | 150 | 3.4 | 4.3 |
| | 165 | 3.6 | 4.5 |
| | 210 | 4.0 | 4.9 |
| | 225 | 4.1 | 5.0 |
| | 240 | 4.1 | 5.0 |

*Appendix Table 4. The calculated H pickup fractions of each validation set and corresponding fitted effective $H_2$ partial pressure from the 10-fold cross validation.*

| Alloy | Corrosion time (day) | $f_{Hvalidation}^{cal}$ (%) | |
|---|---|---|---|
| Zry-2 | 10 | 15.1 | Validation Group 1 |
| | 20 | 19.6 | Fitted $p_{H2}$=2.6×10$^6$ atm |
| | 42 | 27.1 | |
| | 70 | 30.4 | |
| | 105 | 48.3 | |
| GNF-Ziron | 10 | 9.0 | |
| | 20 | 13.4 | |
| | 42 | 20.3 | Validation Group 2 |
| | 70 | 22.5 | Fitted $p_{H2}$=3.0×10$^6$ atm |
| | 105 | 33.4 | |
| VB | 10 | 4.1 | |
| | 20 | 6.5 | |
| | 70 | 12.1 | |
| | 105 | 14.7 | |
| Zry-4(sheet) | 30 | 7.0 | |
| | 45 | 10.7 | Validation Group 3 |
| | 60 | 13.0 | Fitted $p_{H2}$=3.5×10$^6$ atm |
| | 75 | 14.9 | |
| | 90 | 13.6 | |
| | 105 | 13.8 | |
| | 120 | 14.6 | |
| | 135 | 15.3 | |
| | 150 | 16.0 | |
| | 165 | 17.9 | Validation Group 4 |
| Zry-4(tube) | 20 | 6.2 | Fitted $p_{H2}$=3.7×10$^6$ atm |
| | 30 | 8.4 | |
| | 45 | 10.9 | |
| | 60 | 12.9 | |

| | 75 | 14.7 | |
| --- | --- | --- | --- |
| | 90 | 16.0 | |
| | 105 | 17.2 | |
| | 120 | 16.4 | Validation Group 5 |
| | 135 | 14.5 | Fitted $p_{H2}$=3.6×10$^6$ atm |
| | 150 | 15.6 | |
| | 165 | 16.5 | |
| | 210 | 18.6 | |
| | 225 | 19.1 | |
| | 240 | 19.4 | |
| | 255 | 19.7 | |
| ZIRLO(sheet) | 30 | 8.2 | Validation Group 6 |
| | 45 | 9.9 | Fitted $p_{H2}$=3.4×10$^6$ atm |
| | 60 | 10.9 | |
| | 75 | 11.9 | |
| | 90 | 12.5 | |
| | 105 | 12.8 | |
| | 120 | 12.1 | |
| | 135 | 12.7 | |
| | 150 | 13.3 | Validation Group 7 |
| | 165 | 13.5 | Fitted $p_{H2}$=3.4×10$^6$ atm |
| | 210 | 14.2 | |
| | 225 | 13.9 | |
| ZIRLO(tube) | 20 | 4.7 | |
| | 30 | 6.3 | |
| | 45 | 8.0 | |
| | 60 | 9.3 | |
| | 75 | 10.2 | Validation Group 8 |
| | 90 | 11.0 | Fitted $p_{H2}$=3.4×10$^6$ atm |
| | 105 | 11.5 | |
| | 120 | 12.1 | |
| | 135 | 11.1 | |
| | 150 | 11.7 | |
| | 165 | 12.6 | |
| | 210 | 13.9 | |
| | 225 | 13.9 | Validation Group 9 |
| | 240 | 14.0 | Fitted $p_{H2}$=3.3×10$^6$ atm |
| Zr-2.5Nb | 30 | 0.3 | |
| | 45 | 1.2 | |
| | 60 | 1.8 | |
| | 75 | 2.3 | |
| | 90 | 2.6 | |
| | 105 | 3.0 | |
| | 120 | 3.3 | Validation Group 10 |
| | 135 | 3.5 | Fitted $p_{H2}$=3.3×10$^6$ atm |
| | 150 | 3.7 | |
| | 165 | 4.0 | |
| | 210 | 4.4 | |
| | 225 | 4.4 | |
| | 240 | 4.4 | |

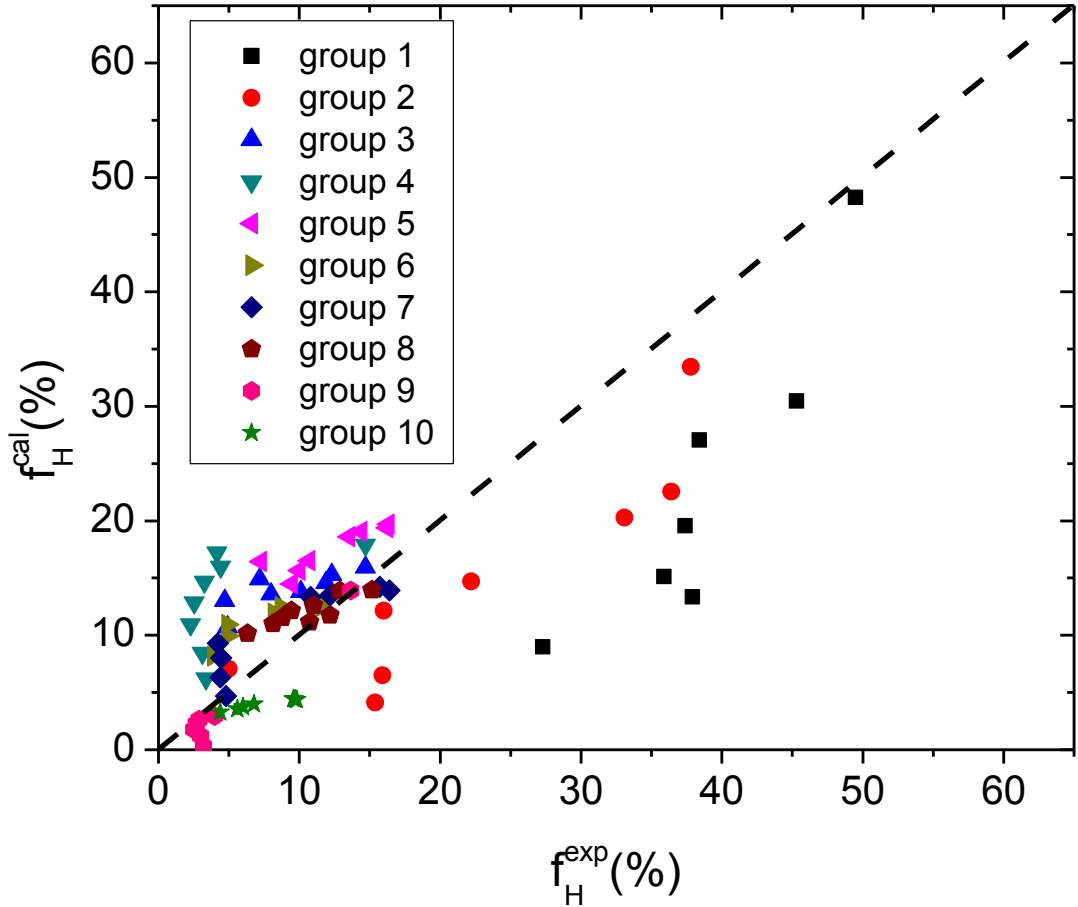

*Appendix Figure 1. Comparison between the experimentally measured $f_H$ and the calculated $f_H$ from the cross validation process. The data points are in the same shape and color if they belong to the same group for the cross validation.*

*Appendix Table 5. Compare the calculated H pickup fractions measured experimentally ($f_H^{exp}$) and H pickup fractions calculated using the model ($f_{H,}^{cal}$) for samples corroded in LiOH solution.*

| Alloy | Corrosion time (day) | $f_H^{exp}$(%) | $f_H^{cal}$(%) |
|---|---|---|---|
| Zry-2 | 1 | 78.7 | 41.0 |
|  | 3 | 87.9 | 55.9 |
|  | 6 | 82.9 | 56.0 |
|  | 9 | 83.9 | 49.7 |
|  | 12 | 80.7 | 43.3 |
| GNF-Ziron | 1 | 18.5 | 24.1 |
|  | 2 | 34.1 | 21.8 |
|  | 3 | 28.2 | 33.3 |
|  | 6 | 29.8 | 33.8 |
|  | 9 | 34.4 | 33.5 |
|    | 12 | 26.8 | 28.5 |
| VB | 1  | 10.8 | 42.8 |
|    | 2  | 4.9  | 48.9 |
|    | 3  | 10.7 | 41.3 |
|    | 6  | 15.1 | 51.1 |
|    | 13 | 15.8 | 45.6 |